\documentclass[
 aps,
 prc,
 amsmath,
 amssymb,
 superscriptaddress,
 reprint,
]{revtex4-2}

\usepackage{graphicx}
\usepackage{dcolumn}
\usepackage{bm}
\usepackage[mathlines]{lineno}

\begin{document}

\preprint{APS/123-QED}

\title[Sample title]{$^{197}$Au($\gamma,\,xn;\,x\,=\,1\thicksim9$) Reaction Cross Section Measurements using \\ Laser-Driven Ultra-Intense $\gamma$-Ray Source}

\author{Di Wu}
\author{Haoyang Lan}
\author{Jianyao Zhang}
\author{Jiaxing Liu}
\author{Huangang Lu}
\author{Jianfeng Lv}
\author{Xuezhi  Wu}
\author{Hui Zhang}
\author{Jie Cai}
\author{Qianyi Ma}
\author{Yuhui Xia}
\author{Zhenan Wang}
\author{Meizhi Wang}
\author{Zhiyan Yang}
\author{Xinlu Xu}
\author{Yixing Geng}
\author{Yanying Zhao}
\author{Chen Lin}
\author{Wenjun Ma}
\affiliation{
State Key Laboratory of Nuclear Physics and Technology,School of Physics, CAPT, Peking University, Beijing 100871, China
}
\affiliation{
Beijing Laser Acceleration Innovation Center, Beijing 101407, China
}

\author{Jinqing Yu}
\affiliation{
School of Physics and Electronics, Hunan University, Changsha 410012, China
}

\author{Haoran Wang}
\author{Fulong Liu}
\author{Chuangye  He}
\author{Bing Guo}
\affiliation{
Department of Nuclear Physics, China Institute of Atomic Energy, Beijing 102413, China
}

\author{Ping Zhu}
\affiliation{
Key Laboratory of High Power Laser and Physics, Chinese Academy of Sciences, Shanghai 201800, China
}

\author{Guoqiang Zhang}
\affiliation{
Shanghai Advanced Research Institute, Chinese Academy of Sciences, Shanghai 201210, China
}

\author{Naiyan Wang}
\affiliation{
Department of Nuclear Physics, China Institute of Atomic Energy, Beijing 102413, China
}

\author{Yugang Ma}
\affiliation{
Key Laboratory of Nuclear Physics and Ion-beam Application (MOE), Institute of Modern Physics, Fudan University, Shanghai 200433, China
}

\author{Xueqing Yan}
 \email{x.yan@pku.edu.cn}
\affiliation{
State Key Laboratory of Nuclear Physics and Technology,School of Physics, CAPT, Peking University, Beijing 100871, China
}
\affiliation{
Beijing Laser Acceleration Innovation Center, Beijing 101407, China
}

\date{\today}

\begin{abstract}
We present a new method for the measurements of photonuclear reaction flux-weighted average cross sections and isomeric ratios using a laser-driven bremsstrahlung $\gamma$-ray source. An ultra-bright ultra-fast 60$\,\thicksim\,$250 MeV bremsstrahlung $\gamma$-ray source was established using the 200 TW laser facility in the Compact Laser Plasma Accelerator Laboratory, Peking University, which could cover the energy range from knocking out neutrons to producing pions. Stable quasi-monoenergetic electron beams were generated via laser wakefield acceleration with a charge of 300$\,\thicksim\,$600 pC per shot. The averaged $\gamma$-ray intensities ($\geqslant$8 MeV) were higher than 10$^{8}$ per shot and the instantaneous intensities can reach above 10$^{19}$ s$^{-1}$ with a duration time about 6.7 ps. $^{65}$Cu($\gamma,\,n$)$^{64}$Cu and $^{27}$Al($\gamma,\,x$)$^{24}$Na reactions were used as $\gamma$-ray flux monitors in the experiments. The flux-weighted average cross sections and isomeric ratios of $^{197}$Au($\gamma,\,xn;\,x\,=\,1\thicksim9$) reactions were analyzed through activation measurements. The results showed good agreement with previous works and proved this method to be accurate. The $^{197}$Au($\gamma,\,xn;\,x\,=\,7\thicksim\,9$) reaction cross sections were first achieved with the highest threshold energy of 71.410 MeV. Theoretical cross sections of TALYS 1.9 were calculated to compare with experiment results. This method offered a unique way of gaining insight into photonuclear reaction research, especially for short-lived isomers which extremely lack experimental data.

\end{abstract}


\maketitle

\section{INTRODUCTION}

Super intense high energy $\gamma$-rays are very useful tools for broad areas. In nuclear physics, high energy $\gamma$-rays for photonuclear reactions and photofission play a crucial role in nuclear astrophysics \cite{bur57,hui60,bla75,arn03}, nuclear reaction mechanisms \cite{arn03}, nuclear structures \cite{hui60,kol98}, and nuclear energy \cite{mac09}. In the field of industrial radiography, laser-driven photon beams can provide high space resolutions and short time resolutions \cite{ben11,wu18,arm19}. Laser-driven intense photon beams are also believed to be a potential method for FLASH radiotherapy, which requires a radiation dose rate higher than 40 Gy/s \cite{mon18,gao22}. For now, high energy $\gamma$-rays were commonly generated by large facilities like electron linear accelerator bremsstrahlung \cite{nai16,vod21}, laser Compton scattering (LCS) \cite{wel09,hor10,hai15,wang22}, in-flight annihilation of monochromatic positrons \cite{car74}, and accelerator-based nuclear reactions \cite{he20}, using which multitudinous achievements in fundamental researches and applications have been achieved.

\begin{figure*}
\includegraphics[scale=0.7]{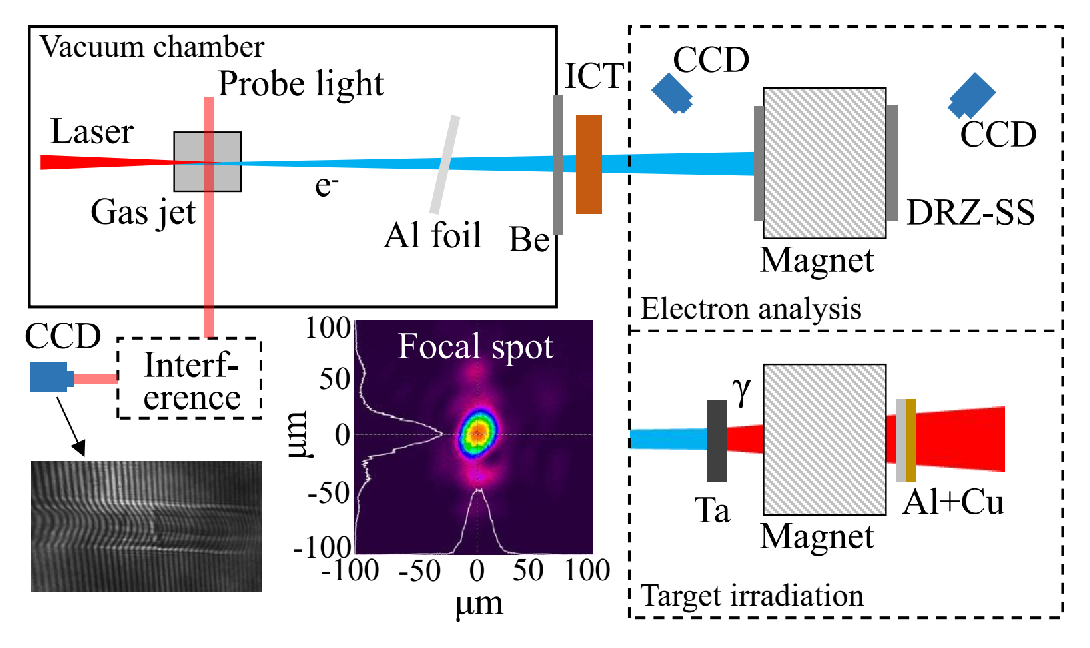}
\caption{\label{fig1} Schematic layout of experimental setup. Typical interference image of the plasma tunnel and typical laser focal spot were also shown.}
\end{figure*}

For photonuclear reactions, their cross sections and isomeric ratios (IR) are very important for nuclear physics studies. The IR of a nuclear reaction depends on the projectile particles and the spins of both target and daughter nuclei, which is used as a good test for nuclear structure theories and nuclear reaction models \cite{van60,hui60,tso00,kol98}. In nuclear astrophysics, there are 30-35 proton-rich nuclei ($p$-nuclei) which can not be produced by neutron-capture processes ($s$-process and $r$-process) \cite{bur57,arn03,arn07}. For those nuclei, a lot of processes, known as $p$-processes, were put forward to explain the $p$-nucleus production mechanism \cite{arn03,sau11,rau13}, which contain a series of photonuclear reactions such as ($\gamma,\,n$), ($\gamma,\,p$), and ($\gamma,\,\alpha$) reactions \cite{rap06}. The experimental measurements of those reactions are very important for nucleosynthesis and stellar models \cite{rap06}. For now, photonuclear reactions are widely studied in their giant dipole resonance (GDR) at $\gamma$-ray energies up to about 30 MeV. In recent years, photodisintegration reactions have attracted a lot of interest and were multiple studies above the GDR \cite{erm10,nai16,vod21}, which are important for the nucleosynthesis, nuclear model test, and applications like accelerator-driven subcritical systems (ADS) \cite{bow03}. But there are very few studies for cross sections of photonuclear reactions with half-life times of several seconds or less due to the lack of suitable $\gamma$-ray sources, which have a significant impact on the nucleosynthesis in the Type II Supernova explosion \cite{arn03,arn07,ray95}. To explore such reactions, ultra-bright ultra-fast $\gamma$-rays are required.

Laser electron accelerations \cite{taj79,mod95,fau04,man04,ged04,puk02,lu06,liu23} have attracted significant interests in nuclear physics such as photonuclear studies \cite{led00,bol21}, photon fission \cite{bol21,boy88,cow00}, production of nuclear medicine radioisotopes \cite{luo16,ma19,sun21} and photon activation analysis \cite{mir21}. Laser-driven bremsstrahlung $\gamma$-ray source, generated by laser wakefield accelerated (LWFA) electrons, can provide a stable ultra-bright, ultra-fast, and energy adjustable photon beam \cite{giu08,fer18,li17,dop16,gun22}, which is suitable for photonuclear reaction cross section and IR measurements. The intensities of laser-driven $\gamma$-rays are several orders of magnitude stronger than other $\gamma$-ray sources, and its duration time can be as short as sub-picosecond, which makes it more favorable for the reaction measurements with very short half-lives. Recently, J. Feng $et. al$ achieved an extreme nuclear isomers product rate of 1.12$\times$10$^{15}$ s$^{-1}$ using LWFA-driven photonuclear reactions \cite{feng23}, proving that laser-driven nuclear reactions can provide a very high contrast ratio for short-lived isomers, which is very beneficial for experimental measurements. But for now, no cross section data has been determined using laser-driven $\gamma$-rays. The biggest reasons are the lack of stability for laser-driven particle beams, and the low average beam current caused by the low repetition frequency, which could make the quantitative analysis very difficult or even impossible. These problems needed to be addressed urgently.

\begin{figure*}
\includegraphics[scale=0.6]{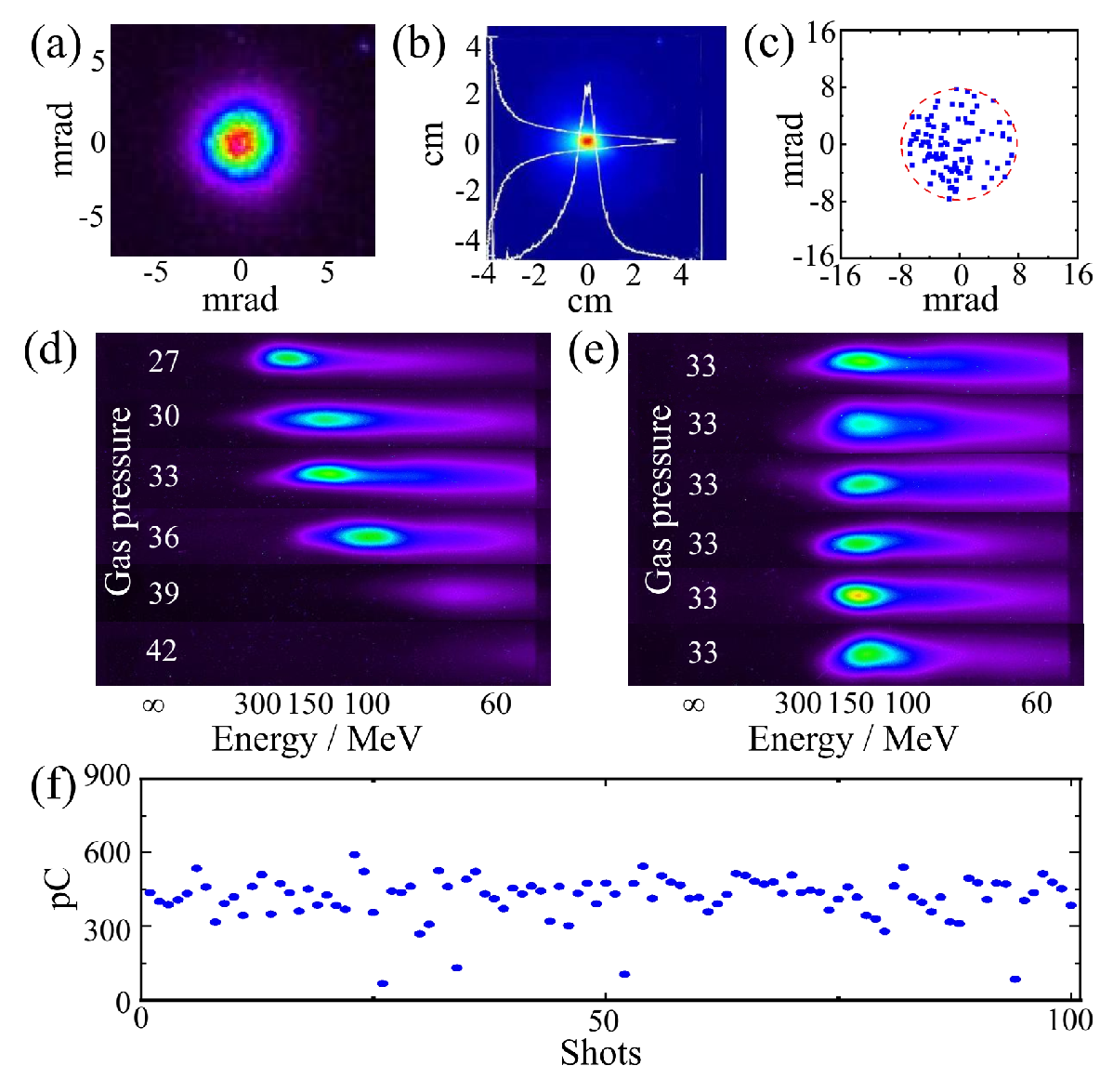}
\caption{\label{fig2} (a) Electron beam at the front surface of the magnetic spectrometer. (b) Electron beam size of 10 continuous shoots at a gas pressure of 33 bar measured with a radiochromic film (RCF). Shadow of the Be window, produced by the X-ray imaging, could also be distinguished on the RCF, which will cause the broadening of the grayscale distribution. (c) Electron directional stability of 100 continuous shots measured at the back surface of magnetic spectrometer at a gas pressure of 33 bar. (d) Electron maximum energies of 60 MeV to 250 MeV at gas pressures of 42 to 27 bar, respectively. (e) Electron energy stability measured at the back surface of the magnetic spectrometer at a gas pressure of 33 bar. (f) Electron charge stability of 100 continuous shots measured with the Turbo-ICT at a gas pressure of 33 bar.}
\end{figure*}

In addition, the accuracy of a new experimental method needs to be tested. Au material has been studied many times in photonuclear experiments due to its single nature isotope, high cross sections, and suitable half-life times of the reaction products. The $^{197}$Au($\gamma,\,xn;\,x\,=\,1\thicksim3$) reactions \cite{ful62,vey70,ber87,har07,kit10,kit11,ito11,pla12,nai16} and the $^{197}$Au($\gamma,\,n$)$^{196m,g}$Au IRs \cite{nai16,sor76,sor78,pal99,gan03,gan04,thi06,vis08,rah10} were measured with different methods, which were often used as high energy $\gamma$-ray monitors. The $^{197}$Au($\gamma,\,xn;\,x\,=\,1\thicksim$6) reaction flux-weighted average cross sections (FACS) were measured using an electron linac \cite{nai16}. The results showed good agreement with previous experiments, and theoretical values using TENDL-2014 \cite{kon12} based on TALYS 1.6 \cite{kon08} were calculated. The measured cross sections of $^{197}$Au($\gamma,\,2n$)$^{195}$Au reaction had a difference of 40\% at maximum compared with theoretical values while other data matched very well, it can be presumed that the TALYS 1.6 code could not describe the $^{197}$Au($\gamma,\,2n$)$^{195}$Au reaction very well. The Au photonuclear reactions were widely used to verify new experimental methods.

In this paper, a laser-driven ultra-intense stable $\gamma$-ray source was established using LWFA electrons with the averaged instantaneous intensities above 10$^{19}$/s. The maximum $\gamma$-ray energy can be adjusted from 60 MeV to 250MeV and the stabilities could reach higher than 90\%. For the first time, the $^{197}$Au($\gamma,\,xn$;\,x\,=\,1$\thicksim$9) reaction FACS and the $^{197}$Au($\gamma,\,n$)$^{196m,g}$Au IRs were measured using laser-driven bremsstrahlung $\gamma$-rays. The $^{197}$Au($\gamma,\,xn$;\,x\,=\,7$\thicksim$9) reaction FACSs were first determined in this work. Theoretical values were calculated and compared with previous experimental results. This method provides a new way to study photonuclear reaction and sheds some light on isomer research with ultra-short half-life times. It is expected to fill the experimental data gap of many nuclear reactions.

\section{LASER-DRIVEN
ULTRA-INTENSE $\gamma$-RAY}

The experiments were performed at the 200 TW laser facility in the Compact Laser Plasma Accelerator (CLAPA) Laboratory, Peking University, which can deliver 30 fs, 5Hz laser pulses in the center wavelength of 800 nm. In this experiment, the laser system was operated in a relatively low-energy state with a delivered energy of about 2.4 J on target. The schematic layout of the experimental setup is shown in Fig.~\ref{fig1}. The laser beam was focused to 21$\,\times\,$23 $\mu$m (1/e$^{2}$ of the maximum intensity) by an f/12.5 off-axis parabolic mirror (OAP). The encircled energy was about 1.2 J corresponding to a normalized vector potential a$_{0}\,\approx\,$1.7, where $a_{0}\,=\,eE_{L}(mc\omega_{0})^{-1}$ is the normalized vector potential of the laser, $E_{L}$ and $\omega_{0}$ are the electric field and the frequency of the laser, $m$ and $e$ are the electron mass and charge, and $c$ is the speed of light in vacuum. The gas jet is a 4$\,\times\,$1 mm rectangular supersonic nozzle, which can provide a flat density distribution. Pure helium gas is used and the operating pressure can be adjusted from 0 bar to 45 bar. Electrons are accelerated in LWFA with a monoenergetic spectrum \cite{puk04}. The electron energy spectra are analyzed by a 0.8 T magnetic spectrometer 100 cm away from the gas jet, fluorescent screens (Gd$_{2}$O$_{2}$S:Tb), and CCD cameras are used to record electron signals.

Electron parameters were analyzed before the irradiation experiment and shown in Fig.~\ref{fig2}. The maximum electron energies can be adjusted in the range of 60 MeV to 250 MeV by changing the gas pressures from 42 to 27 bar. Gas pressures at 33, 36, and 39 bar were used in this work due to suitable electron energies and charges. When operating at these conditions, the laser pulse self-focused further to a higher a$_{0}$, and then defocused and generated an expanding bubble where plenty of electrons were self-injected and accelerated. To determine the stability of the electron parameters, 100 continuous shots were recorded at each gas pressure. The divergence angle was about 4.3 mrad, the pointing stability was better than 5.2 mrad, and the averaged center energies were 135$\,\pm\,$20, 103$\,\pm\,$14 and 78$\,\pm\,$10 MeV, respectively. The energy uncertainties were given by the full width at half maximum (FWHM) of the averaged electron spectrum.  The center energy of several shots (5$\,\thicksim\,$9 shots in 100 shots) might fall out of the uncertainty range, which might lead to deviation of the $\gamma$-ray density determinations and were correspondingly considered in the following calculations. The electron charges were 300$\,\thicksim\,$600 pC measured with a Turbo Integrating Current Transformer (Turbo-ICT) \cite{nak16} out of a 40 $\mu$m Al foil and a 60 $\mu$m Be window, which were used to block the scattered light and to seal the vacuum.

\begin{figure}
\includegraphics[scale=0.35]{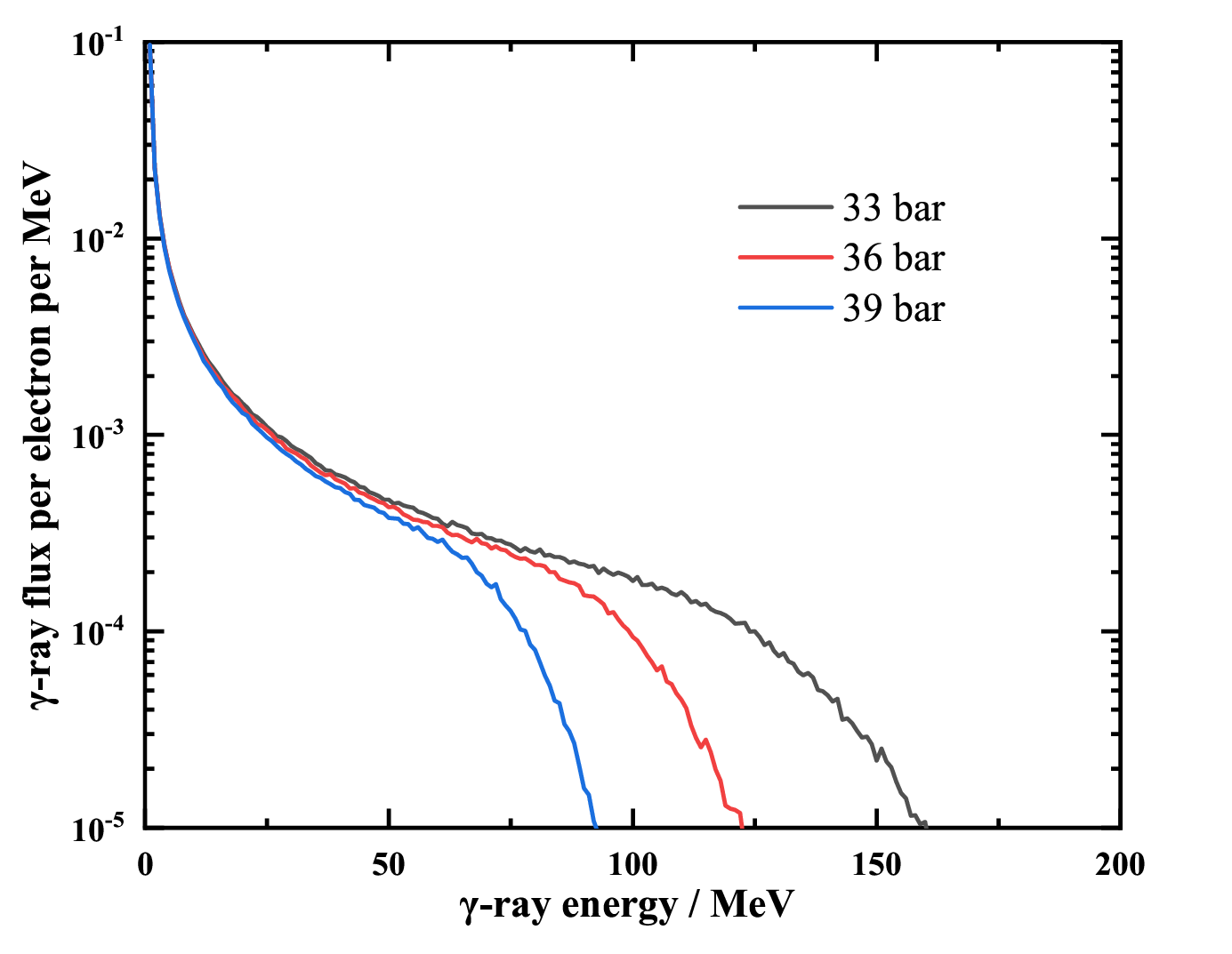}
\caption{\label{fig3} Bremsstrahlung $\gamma$-ray spectra per electron at different gas pressures calculated by GEANT4 simulation using the averaged electron energy spectra of 100 continuous shots.}
\end{figure}

A Ta (99.9\%) disk with a size of 2$\,\times\,$2 cm and a thickness of 2 mm, placed at 5 cm before the front surface of the magnetic spectrometer, was used as a high-Z bremsstrahlung converter. The remaining electrons were deflected away by the magnet behind the Ta disk. The bremsstrahlung $\gamma$-ray spectra per electron (shown in Fig.~\ref{fig3}) were obtained by the average electron spectra of 100 continuous shots and GEANT 4 simulations \cite{ago03,all06}. And the $\gamma$-ray intensities (shown in Tab.~\ref{tab1}) were given by the electron charges measured by the Turbo-ICT.

\begin{table}
\caption{\label{tab1}$\gamma$-ray intensities ($\geqslant$8 MeV) at different energies using a converter of 2 mm Ta.}
\begin{ruledtabular}
\begin{tabular}{ccc}
Electron energy & $\gamma$-ray intensity & Instantaneous intensity\\
MeV & per shoot & s$^{-1}$\\
\hline
78$\,\pm\,$10 & (1.60$\,\pm\,$0.14)$\times$10$^{8}$ & (2.39$\,\pm\,$0.20)$\times$10$^{19}$\\
103$\,\pm\,$14 & (1.76$\,\pm\,$0.13)$\times$10$^{8}$ & (2.63$\,\pm\,$0.19)$\times$10$^{19}$\\
135$\,\pm\,$20 & (1.84$\,\pm\,$0.11)$\times$10$^{8}$ & (2.74$\,\pm\,$0.17)$\times$10$^{19}$\\
\end{tabular}
\end{ruledtabular}
\end{table}

The accurate $\gamma$-ray intensities are very important in photonuclear reaction studies. Therefore, the $^{65}$Cu($\gamma,\,n$)$^{64}$Cu reaction with a threshold energy at 9.91 MeV and $^{27}$Al($\gamma,\,2pn$)$^{24}$Na reaction with a threshold energy at 31.45 MeV were used as $\gamma$-ray flux monitors in our experiments. Both $^{27}$Al (99.99\%) targets and $^{nat}$Cu (99.99\%, 30.85\% $^{65}$Cu and 69.15\% $^{63}$Cu) targets had a size of 20$\,\times\,$20$\,\times\,$1 mm. The repetition frequency was set at 0.25 Hz to keep the vacuum at acceptable levels and the irradiation time was 1 h for each electron energy. For each shot at a certain energy, the reaction yield can be given by

\begin{equation}
Y_{i}(E) =
N_{s}N_{i}^{e}\int_{E_{thr}}^{E_{max}} \sigma(E)\varphi(E) dE,
\label{eq1}
\end{equation}

where $i$ is number of shots, $N_{s}$ is the the target nuclear number in per unit area, $N_{i}^{e}$ is the number of electrons measured by the Turbo-ICT, $E_{thr}$ is the reaction threshold, $E_{max}$ is the maximum energy of the $\gamma$-ray, $\sigma(E)$ is the energy-dependent reaction cross section calculated using TALYS1.9 code \cite{alh22}, and $\varphi(E)$ is the bremsstrahlung $\gamma$-ray flux per electron. For $^{27}$Al($\gamma,\,2pn$)$^{24}$Na reaction reaction cross sections, a gain factor of 2.4 was used in the calculation due to previous experiment results \cite{vod21Al,dei22}.

Different with electron linac bremsstrahlung or LCS $\gamma$-rays, laser-driven $\gamma$-rays had low repetition frequencies and the $\gamma$-ray intensities of each shot could differ a lot, therefore, it can not be treated as a continuous beam. The radioactive nuclide number of a reaction product at the end of irradiation was determined by

\begin{equation}
N_{0}(E) =
\sum_{i=1}^{n} Y_{i}(E)exp({-\lambda\frac{n-i}{f}}),
\label{eq2}
\end{equation}

where $n$ is the total shots, $\lambda$ is the decay constant of the reaction product, $f$ is the repetition frequency.

\begin{table*}
\caption{\label{tab2}Relative nuclear spectroscopic data of the radioactive nuclei from the $^{197}$Au photonuclear reactions. The energies and branch ratios of $^{189}$Au to $^{196}$Au were given by NuDat 3.0 \cite{nudat} while ones of $^{188}$Au were given by Ref.\cite{sin90}.}
\begin{ruledtabular}
\begin{tabular}{cccccccc}
 Nucleus\footnotemark[1] & Half-life & Decay mode & $\gamma$-ray energy / keV & $\gamma$-ray branch & Threshold / MeV\\
 \hline
 $^{196g}$Au & 6.1669 d & $\beta^{-}$ 7.0\% & 355.73\footnotemark[2] & 87\% & 8.072\\
 &  & $\epsilon$ 93.0\% & 333.03\footnotemark[2] & 22.9\% &\\
 &  &  & 426.10\footnotemark[2] & 6.6\% &\\
 $^{196m1}$Au & 8.1 s & IT 100\% & 84.66 & 0.305\% & 8.157\\
 $^{196m2}$Au & 9.6 h & IT 100\% & 147.81\footnotemark[2] & 43.5\% & 8.668\\
 &  &  & 188.27 & 30.0\% & \\
 $^{195}$Au & 186.01 d & $\epsilon$ 100\% & 98.857\footnotemark[2] & 11.21\% & 14.715\\
 $^{194}$Au & 38.02 h & $\epsilon$ 100\%  & 293.549 & 10.9\% & 23.142\\
 &  &  & 328.470\footnotemark[2] & 62.8\% & \\
 &  &  & 511.0 & 3.77\% & \\
 &  &  & 1468.904 & 6.80\% & \\
 &  &  & 2043.719 & 3.74\% & \\
 $^{193}$Au & 17.65 h & $\epsilon$ 100\% & 173.52 & 2.8\% & 30.020\\
 &  &  & 186.17 & 9.7\% & \\
 &  &  & 255.57\footnotemark[2] & 6.5\% & \\
 $^{192}$Au & 4.94 h & $\epsilon$ 100\% & 295.96 & 23\% & 38.724\\
 &  &  & 308.46 & 3.5\% & \\
 &  &  & 316.57\footnotemark[2] & 59\% & \\
 &  &  & 511.0 & 11\% & \\
 &  &  & 612.46 & 4.4\% & \\
 &  &  & 2237.3 & 4.7\% & \\
 $^{191}$Au & 3.18 h & $\epsilon$ 100\% & 277.86 & 6.4\% & 45.770\\
 &  &  & 283.90 & 5.9\% & \\
 &  &  & 586.44\footnotemark[2] & 15\% & \\
 $^{190}$Au & 42.8 m & $\epsilon$ 100\% & 295.82\footnotemark[2] & 90\% & 50.805\\
 &  &  & 301.82\footnotemark[2] & 30\% & \\
 &  &  & 511.0 & 16.0\% & \\
 &  &  & 597.68 & 12\% & \\
 $^{189g}$Au & 28.7 m & $\epsilon$ 100\% & - & - & 62.128\\
 $^{189m}$Au & 4.59 m & $\epsilon$ 100\% & 166.40\footnotemark[2] & 59\% & 62.375\\
 &  &  & 321.1 & 11.9\% & \\
 &  &  & 511.0 & 20.0\% & \\
 $^{188}$Au & 8.84 m & $\epsilon$ 100\% & 265.63\footnotemark[2] & 77.5\% & 71.410\\
 &  &  & 330.76 & 3.86\% & \\
 &  &  & 340.04 & 18.53\% & \\
\end{tabular}
\footnotetext[1]{$^{188m,190m-195m}$Au are not shown in this table due to their short half-live times, all those isomers will decay to their ground states.}
\footnotetext[2]{Those $\gamma$-rays were used to determine the reaction cross sections.}
\end{ruledtabular}
\end{table*}

Two HPGe detectors with relative efficiencies of 40\% were used to measure the target activities 10 min after the irradiation, all detectors were shielded in Pb brick. The detector efficiencies were calibrated with a $^{152}$Eu source and a $^{60}$Co source, and finally determined by GEANT4 simulations \cite{ago03,all06}. The measured activities of $^{64}$Cu and $^{24}$Na matched well with the calculated values within the uncertainties, which confirmed the accuracy of the $\gamma$-ray intensities.

\section{$^{197}$AU photodisintegration REACTION MEASUREMENTS}

\begin{figure*}
\includegraphics[scale=0.95]{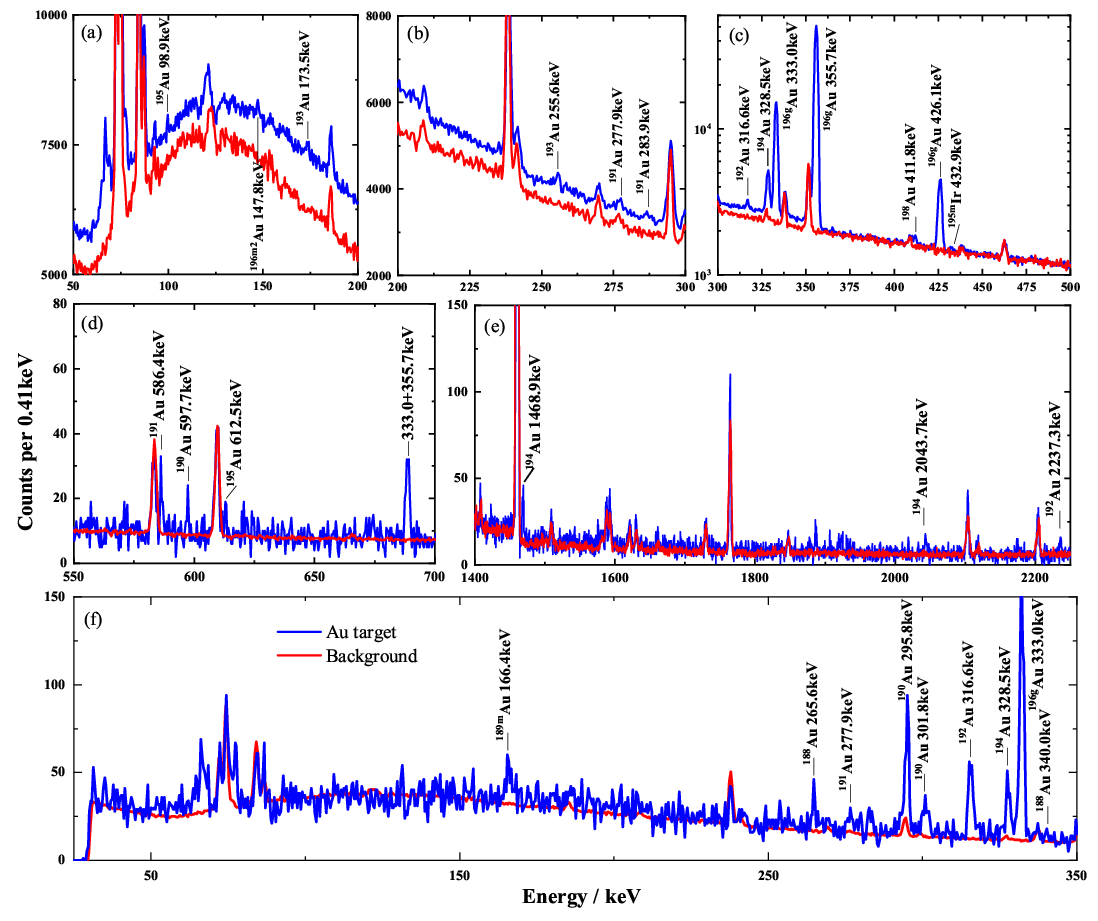}
\caption{\label{fig4} Typical spectra of Au target activation signals at a gas pressure of 36 bar. (a,b,c,e) were the same spectrum with a long measuring time while (d) had a relatively short measuring time to show the decay of $^{190}$Au. (e) showed the decay of $^{189m}$Au and $^{188}$Au with a very short measuring time.}
\end{figure*}

The $^{197}$Au($\gamma,\,xn;\,x\,=\,1\thicksim9$) photodisintegration reactions were measured using $^{197}$Au (99.99\%) targets with the same sizes of Al and Cu targets. The repetition frequency was also set at 0.25 Hz and the irradiation time was still 1 h for each electron energy. Al and Cu targets, putting after the Au targets, were also used as $\gamma$-ray flux monitors. The relative nuclear decay data of the radioactive nuclei from the reaction product $^{188-196}$Au were listed in Table~\ref{tab2}. A typical activation spectrum of the Au target measured by the HPGe detector was shown in Fig.~\ref{fig4}, compared with a living-time normalized background spectrum using a blank Au target. The decay characteristic $\gamma$-ray signals of $^{188g,190-196g}$Au and $^{189m,196m2}$Au were clearly distinguished and were reconfirmed by the half-life times.

After the radioactive nuclide numbers of each products were determined, the FACS can be given by

\begin{equation}
\sigma_{FA}(E) =
\frac
{N_{0}}
{N_{s}\int_{E_{thr}}^{E_{max}} \varphi(E) dE
\cdot\sum_{i=1}^{n}N_{i}^{e}exp({-\lambda\frac{n-i}{f}})}.
\label{eq3}
\end{equation}

$^{196}$Au has two isomeric states: $^{196m1}$Au and $^{196m2}$Au. $^{196m1}$Au has a half-life time of 8.1 s and could not be detected in this offline method. $^{196m2}$Au has a half-life time of 9.6 h, which can be built up to sufficient intensities. Both isomer states will decay to the ground state. Therefore, the IRs of $^{197}$Au($\gamma,\,n$)$^{196m,g}$Au reaction is given by

\begin{equation}
IR =
\frac
{\sigma^{m2}_{FA}(E)}
{\sigma^{m1}_{FA}(E)+\sigma^{g}_{FA}(E)},
\label{eq4}
\end{equation}
where $m$ and $g$ stand for the parameters of isomeric state and ground state, respectively.

Experimental FACSs and IRs of $^{197}$Au($\gamma,\,xn$;\,x\,=\,1$\thicksim$9) reactions were listed in Tab.~\ref{tab3}. The statistical uncertainties in this experiment range from 0.22\% to 8.2\% for different decay $\gamma$-rays, and the systematic uncertainties include the electron instabilities at each energy, the electron charge uncertainties measured by the Turbo-ICT, the calibration uncertainties of detection efficiencies, the decay branch uncertainties of each isotope, and the corrections of activation targets.

\begin{table*}
\caption{\label{tab3} The experimental FACSs of $^{197}$Au($\gamma,\,xn$;\,x\,=\,1$\thicksim$9) reactions determined by this work and theoretical values calculated by TALYS 1.9.}
\begin{ruledtabular}
\begin{tabular}{cccc}
Reaction & Center energy of electrons & Experimental cross section & TALYS 1.9\footnotemark[1] \\
 & MeV & mb & mb \\
\hline
$^{197}$Au($\gamma,\,n$)$^{196g+m1}$Au & 78$\,\pm\,$10 & 95.6$\,\pm\,$8.3 & 92.6 \\
 & 103$\,\pm\,$14 & 90.2$\,\pm\,$8.2 & 87.1 \\
 & 135$\,\pm\,$20 & 79.5$\,\pm\,$7.4 & 76.5 \\
$^{197}$Au($\gamma,\,n$)$^{196m2}$Au & 78$\,\pm\,$10 & 0.0549$\,\pm\,$0.0054 & 0.0512 \\
 & 103$\,\pm\,$14 & 0.0538$\,\pm\,$0.0053 & 0.0487 \\
 & 135$\,\pm\,$20 & 0.0492$\,\pm\,$0.0052 & 0.0461 \\
 $^{197}$Au($\gamma,\,2n$)$^{195}$Au & 78$\,\pm\,$10 & 25.0$\,\pm\,$2.4 & 31.5 \\
 & 103$\,\pm\,$14 & 21.7$\,\pm\,$2.2 & 27.5 \\
 & 135$\,\pm\,$20 & 17.9$\,\pm\,$1.9 & 24.6 \\
 $^{197}$Au($\gamma,\,3n$)$^{194}$Au & 78$\,\pm\,$10 & 6.23$\,\pm\,$0.60 & 6.2 \\
 & 103$\,\pm\,$14 & 5.35$\,\pm\,$0.54 & 5.33 \\
 & 135$\,\pm\,$20 & 5.01$\,\pm\,$0.52 & 4.72 \\
 $^{197}$Au($\gamma,\,4n$)$^{193}$Au & 78$\,\pm\,$10 & 4.92$\,\pm\,$0.43 & 4.75 \\
 & 103$\,\pm\,$14 & 4.21$\,\pm\,$0.38 & 4.03 \\
 & 135$\,\pm\,$20 & 3.62$\,\pm\,$0.34 & 3.51 \\
 $^{197}$Au($\gamma,\,5n$)$^{192}$Au & 78$\,\pm\,$10 & 3.42$\,\pm\,$0.32 & 3.08 \\
 & 103$\,\pm\,$14 & 2.79$\,\pm\,$0.27 & 2.53 \\
 & 135$\,\pm\,$20 & 2.36$\,\pm\,$0.24 & 2.17 \\
 $^{197}$Au($\gamma,\,6n$)$^{191}$Au & 78$\,\pm\,$10 & 3.02$\,\pm\,$0.28 & 2.81 \\
 & 103$\,\pm\,$14 & 2.45$\,\pm\,$0.23 & 2.25 \\
 & 135$\,\pm\,$20 & 1.92$\,\pm\,$0.19 & 1.88 \\
 $^{197}$Au($\gamma,\,7n$)$^{190}$Au & 78$\,\pm\,$10 & 1.81$\,\pm\,$0.21 & 1.81 \\
 & 103$\,\pm\,$14 & 1.78$\,\pm\,$0.22 & 1.57 \\
 & 135$\,\pm\,$20 & 1.44$\,\pm\,$0.19 & 1.24 \\
 $^{197}$Au($\gamma,\,8n$)$^{189m}$Au & 103$\,\pm\,$14 & 1.13$\,\pm\,$0.14 & 1.03 \\
 & 135$\,\pm\,$20 & 0.88$\,\pm\,$0.11 & 0.78 \\
 $^{197}$Au($\gamma,\,9n$)$^{188}$Au & 103$\,\pm\,$14 & 1.06$\,\pm\,$0.17 & 1.13 \\
 & 135$\,\pm\,$20 & 0.85$\,\pm\,$0.12 & 0.82 \\
\end{tabular}
\footnotetext[1]{The theoretical values were calculated using monoenergetic electron beams with the energies of 78, 103 and 135 MeV.}
\end{ruledtabular}
\end{table*}

Theoretical values (shown in Tab.~\ref{tab3}) of FACSs and IRs for $^{197}$Au photonuclear reactions were calculated by GEANT 4 simulation using the data from TENDL-2019 library \cite{kon19} based on TALYS 1.9 code \cite{kon12,alh22}. Theoretical FACSs were defined as

\begin{equation}
\sigma_{FA}(E) =
\frac
{\int_{E_{thr}}^{E_{max}} \sigma(E)\varphi(E) dE}
{\int_{E_{thr}}^{E_{max}} \varphi(E) dE}.
\label{eq5}
\end{equation}

Monoenergetic electron energies were used in the calculations. Theoretical IRs were also obtained by Eq.~\ref{eq4}.

\begin{figure}
\includegraphics[scale=0.35]{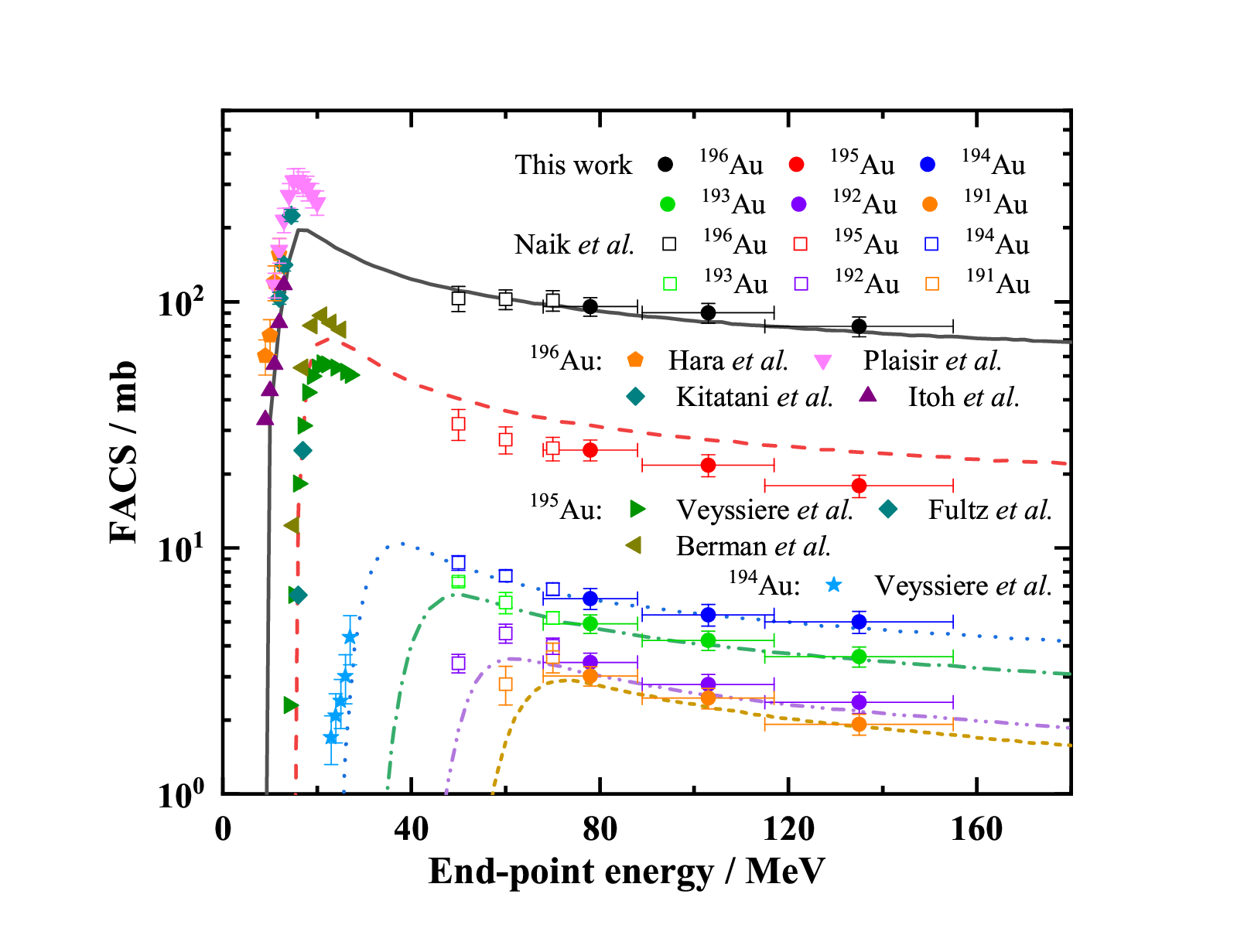}
\caption{\label{fig5} Experimental results of $^{197}$Au($\gamma,\,xn$;\,x\,=\,1$\thicksim$6) reaction FACSs, compared with previous works \cite{ful62,vey70,ber87,har07,kit10,kit11,ito11,pla12,nai16} and TALYS 1.9 calculations. Lines are the flux-weighted average cross sections of $^{197}$Au($\gamma,\,n$)$^{196}$Au to $^{197}$Au($\gamma,\,6n$)$^{191}$Au reaction from the top to the bottom, respectively. The bremsstrahlung end-point energies of our experimental cross sections were given as the averaged center electron energies.}
\end{figure}

As shown in Fig.~\ref{fig5}, our experimental FACSs of $^{197}$Au($\gamma,\,xn$;\,x\,=\,1$\thicksim$6) reactions matched with previous works \cite{ful62,vey70,ber87,har07,kit10,kit11,ito11,pla12,nai16} and most theoretical values except the $^{197}$Au($\gamma,\,2n$)$^{195}$Au reaction. Theoretical values given by TALYS 1.9 are higher than our experiment results by about 37\% at maximum, which is similar to the results of H. Naik $et. al$ \cite{nai16}. TALYS 1.9 code still can not describe the $^{197}$Au($\gamma,\,2n$)$^{195}$Au reaction very well.

\begin{figure}
\includegraphics[scale=0.35]{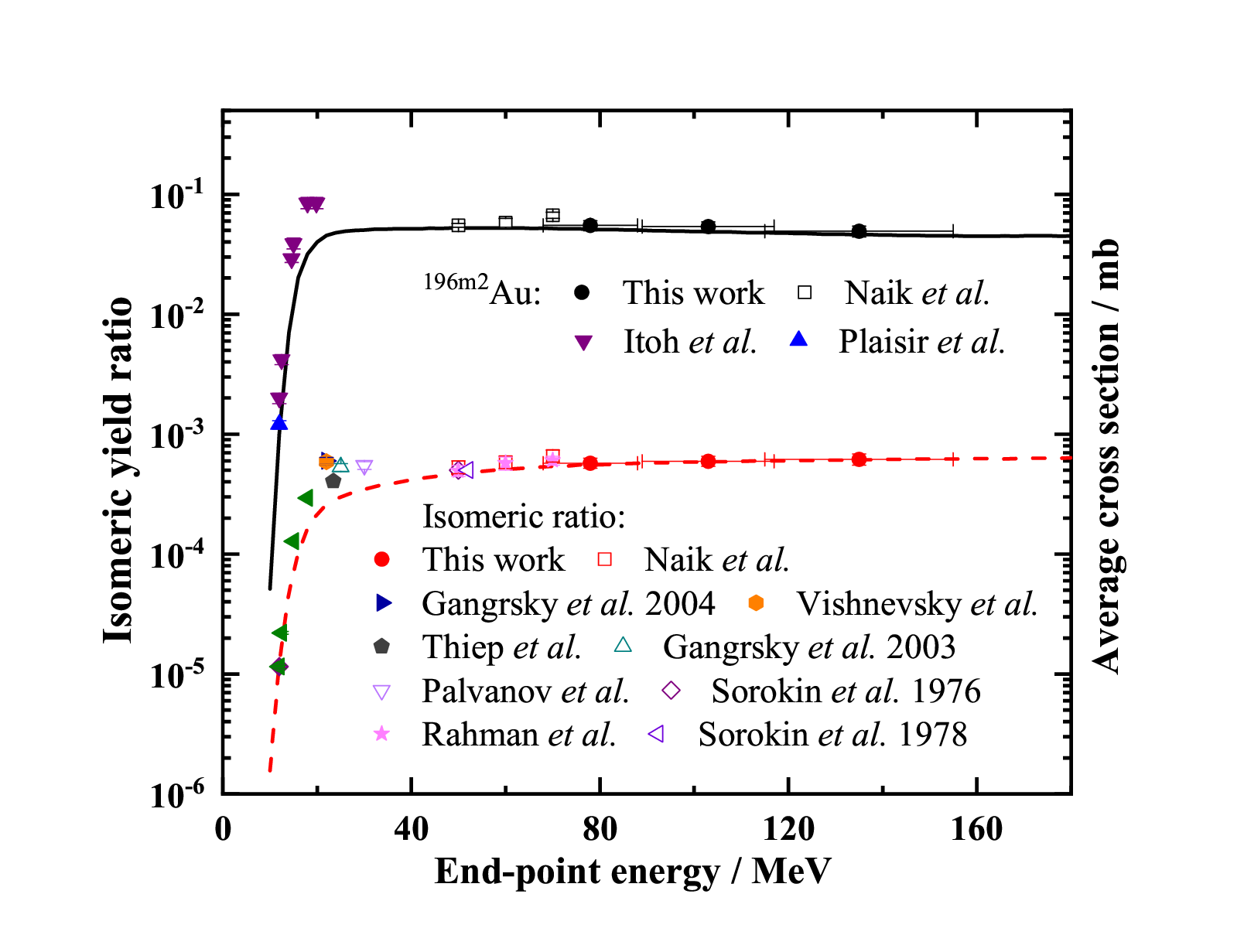}
\caption{\label{fig6} Experimental results of $^{197}$Au($\gamma,\,n$)$^{196m2}$Au reaction FACSs and $^{197}$Au($\gamma,\,n$)$^{196m,g}$Au IRs, compared with previous works \cite{ito11,nai16,sor76,sor78,pal99,gan03,gan04,thi06,vis08,rah10} and TALYS 1.9 calculations. Solid line is the $^{197}$Au($\gamma,\,n$)$^{196m2}$Au reaction FACSs, and dashed line is the $^{197}$Au($\gamma,\,n$)$^{196m,g}$Au reaction IRs.}
\end{figure}

Fig.~\ref{fig6} showed the experimental results of $^{197}$Au($\gamma,\,n$)$^{196m2}$Au reaction FACSs and $^{197}$Au($\gamma,\,n$)$^{196m,g}$Au IRs, both matched well with previous works \cite{ito11,nai16,sor76,sor78,pal99,gan03,gan04,thi06,vis08,rah10} and TALYS 1.9 calculations. The results of $^{197}$Au($\gamma,\,xn$;\,x\,=\,1$\thicksim$6) reaction proved the correctness of the experiment method and data analysis in our work.

\begin{figure}
\includegraphics[scale=0.35]{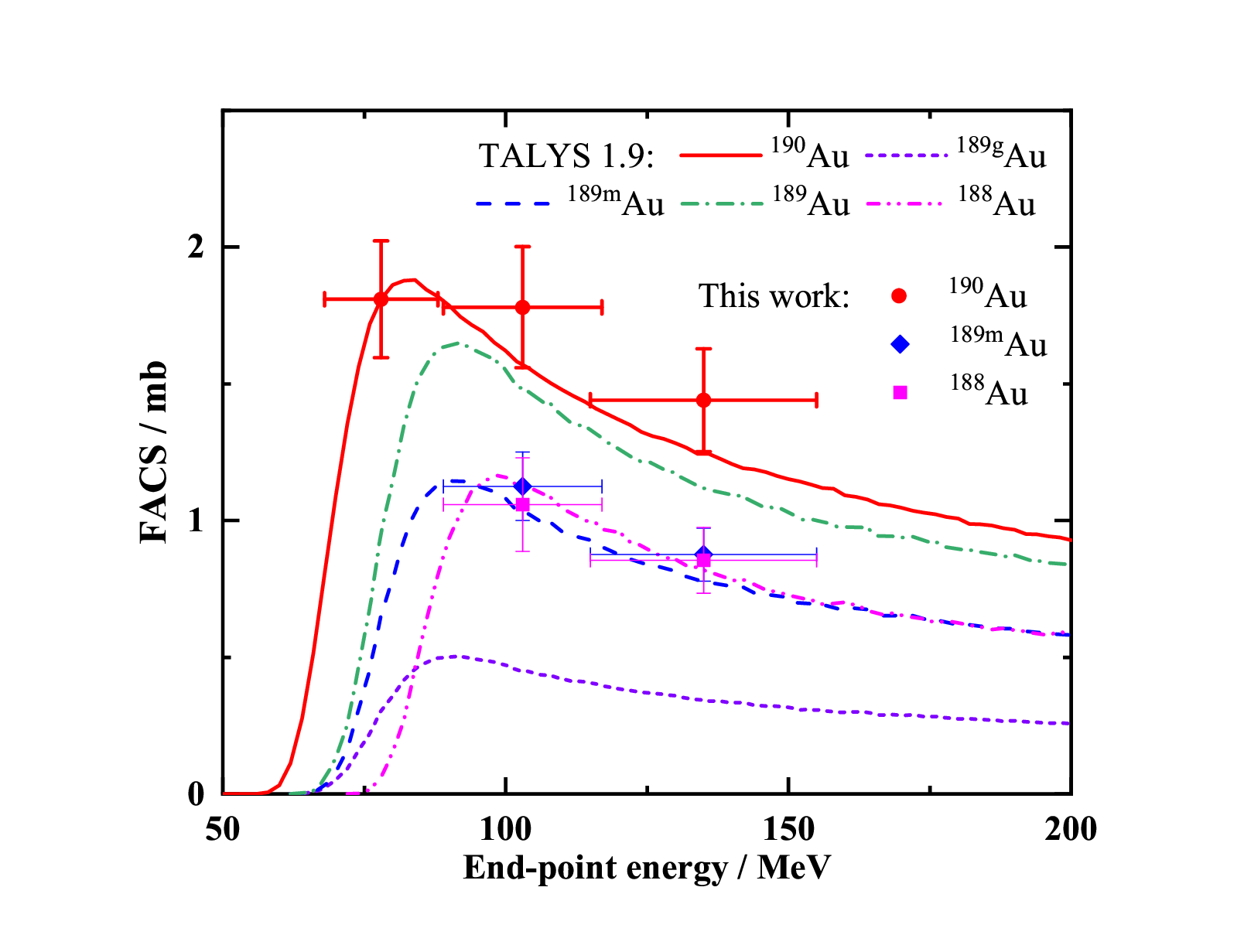}
\caption{\label{fig7} Experimental results of $^{197}$Au($\gamma,\,xn;\,x\,=\,7\thicksim$9) reaction FACSs, compared with TALYS 1.9 calculations.}
\end{figure}

For the $^{197}$Au($\gamma,\,xn$;\,x\,=\,7$\thicksim$9) reactions, the reaction thresholds are relatively high and the energy broadening of LWFA electrons need to be considered. When electron energies are close to the reaction thresholds, reaction cross sections change dramatically with the energies, which will lead to relatively low experimental results due to the low-energy electrons having much lower contributions to the reaction yields, such as the FACSs of the $^{197}$Au($\gamma,\,7n$)$^{190}$Au reaction at an electron energy of 78$\,\pm\,$10 MeV and the $^{197}$Au($\gamma,\,9n$)$^{188}$Au reaction at an electron energy of 103$\,\pm\,$14 MeV. For other experiment results, theoretical values are slightly lower than experiment results for $^{197}$Au($\gamma,\,7n$)$^{190}$Au reaction and $^{197}$Au($\gamma,\,8n$)$^{189m}$Au reaction, and matched within the experiment uncertainty for $^{197}$Au($\gamma,\,9n$)$^{188}$Au reaction. These cross section data were first achieved in this work, which will provide new references for the study of nuclear structure and photon induced evaporation process.

\section {CONCLUSION}

In summary, we present a new method for the studies of photonuclear reactions using a laser-driven $\gamma$-ray source, which has an ultra-brightness of 10$^{19}$ s$^{-1}$ (10$^{8}$ per shoot) and an ultra-short duration time of less than 10 ps. The $^{197}$Au($\gamma,\,xn$;\,x\,=\,1$\thicksim$9) reactions were measured and compared with previous experimental works and TALYS 1.9 calculations, which proved this new method to be accurate. The $^{197}$Au($\gamma,\,xn$;\,x\,=\,7$\thicksim$9) reaction FACSs were first measured in this work and provided more experiment data for related research. The results can be also used as monitors in similar experiments with high $\gamma$-ray energies.

The nuclear structure and nuclear reaction mechanisms of $^{197}$Au were well restrained due to abundant previous experiment works, but for most nuclei, more experimental data are still needed. This new method offered an effective way to study photonuclear reaction cross sections from single nucleon knock-out reactions to photon induced evaporation process, even for making pions. The ultra-short duration times of laser-driven $\gamma$-rays can build up extreme product rates higher than 10$^{15}$ s$^{-1}$ for short-lived nuclear isomers, which barely had experimental data. Those short-lived nuclear isomers have proved to be of great interest in both fundamental physics studies and have great application prospects such as nuclear laser \cite{tka11} or nuclear clock \cite{cam12,sei19}. This work offered a unique way of gaining insight into related experimental research.

\section{ACKNOWLEDGMENTS}

The authors thank the staff of the 200 TW laser of CLAPA laboratory for the smooth operation of the machine. We thank Professor Wenqing Shen, Professor Baozhen Zhao, and Professor Chong Lv for their helpful comments and suggestions.

This work was supported by the National Natural Science Foundation of China (Grants No.12305266, 11921006, 12125509), the National Grand Instrument Project (No. 2019YFF01014400), Beijing Outstanding Young Scientists Program, and the Open Foundation of Key Laboratory of High Power Laser and Physics, Chinese Academy of Sciences (No. SGKF202104).

\bibliography{apssamp}

\end{document}